\documentstyle{article}\begin{document}
\title{Semi-classical Einstein equations:descend to the ground state }
\author{ Zbigniew Haba\\
Institute of Theoretical Physics, University of Wroclaw,\\ 50-204
Wroclaw, Plac Maxa Borna 9, Poland\\
email:zbigniew.haba@uwr.edu.pl}\maketitle
\begin{abstract} The time-dependent cosmological term arises from
the energy-momentum tensor calculated in a state different from
the ground state. We discuss the expectation value of the
energy-momentum tensor on the rhs of Einstein equations in various
(approximate) quantum pure as well as mixed states. We apply the
classical slow-roll field evolution as well as the Starobinsky and
warm inflation stochastic equations in order to calculate the
expectation value. We show that in the state concentrated at the
local maximum  of the double-well potential the expectation value
is decreasing exponentially. We confirm the descend of the
expectation value in the stochastic inflation model. We calculate
the cosmological constant $\Lambda$ at large time as the
expectation value of the energy density with respect to the
stationary probability distribution. We show that $\Lambda\simeq
\gamma^{\frac{4}{3}}$ where $\gamma$ is the thermal dissipation
rate.
\end{abstract}
\section{Introduction} \label{sec.1}
In the description of inflation, it is understood that at the
early stages of the universe evolution quantum physics is
relevant. It is not clear how to take quantum effects into account
because the Einstein equations are classical. For later evolution
leading to the structure formation, the quantization of gravity
and the scalar field in a linear approximation around the
homogeneous classical solution leads to satisfactory results
\cite{mukhanov,flu1,flu2,brand}. The standard approach
\cite{brand,guth} to inflation driven by the scalar field is in
fact semi-classical. It is assumed that the classical scalar field
evolution arises in a quantum field theory in the classical limit
\cite{guth-pi}. The quantum fluctuations as a result of inflation
lose their quantum character and behave as classical fluctuations
\cite{st1,st2}. The quantum statistical mechanics is applied in
order to use the result that the quantum system evolves to its
ground state. For a typical double-well potential, the quadratic
approximation has a negative mass square term. Such a quadratic
Lagrangian has a local maximum at $\phi=0$. The classical system
starting close to $\phi=0$ will quickly evolve to $\phi\neq 0$.
The same happens in quantum mechanics \cite{guth-pi}. The Gaussian
state localized at $\phi=0$ evolves exponentially fast to a
localization at $\phi\neq 0$. In another description
\cite{col1,col2,patra}, one describes the state evolution from one
side of the barrier $\phi<0$ to another side with
 $\phi> 0$ as
the barrier penetration. Then, one can calculate the resulting
wave function \cite{patra} by semi-classical imaginary time
functional integral. By Einstein equations, the universe evolution
is driven by the expectation value of the energy-momentum. The
different speed of the expansion of the universe can be explained
by the time dependence of this expectation value (time dependent
cosmological term). In other words, the time dependence of the
cosmological term can be associated with the evolution from the
false vacuum to the ground state \cite{lima,review,weinberg}. From
CMB observations, we know that the universe was once (before
reionization) in the thermal state. It may be that such a thermal
state appeared already at an earlier stage of universe evolution
\cite{bererafang,hindu}. The barrier penetration takes place also
in thermal systems \cite{warmtun}. For a quantum description, we
need the quantum mechanics in a thermal state.

In this paper, we consider a method of the description of the
vacuum decay based on the slow-roll and diffusion approximations
to the quantum evolution in an expanding universe
\cite{starobinsky,berrerarev}. This is a long wave approximation
which neglects the second-order derivatives in equations of motion
and first-order derivatives in the energy-momentum tensor. The
noise in the stochastic equation consists of two independent
parts: the quantum noise of Starobinsky \cite{starobinsky} and
Vilenkin \cite{vilenkin} and the thermal noise describing thermal
fluctuations \cite{berrerarev,ramos,int,habathermal}.
 We
consider Einstein equations with the expectation value of the
energy-momentum tensor in a quantum state $\psi$ on the rhs of
these equations. There is an unspecified problem of the choice of
$\psi$ and a difficult task to determine its time evolution on a
non-perturbative level (to treat a transition between states which
are not related by perturbation theory).
 We wish to investigate the time evolution of the expectation value of the energy density
 in a state simulating the false vacuum and for a comparison in a state close to the true ground state. The aim is to study the behavior of the
 cosmological term during the transition from the false vacuum to the ground state.
We choose as a test state first the state localized near the local
maximum of the potential and subsequently a state localized close
to the minimum of the potential. The time evolution of the states
is equivalent to the time evolution of the fields defining the
energy-momentum. First, we obtain this time evolution in a
classical and slow-roll approximation. Then, the expectation value
is calculated. It is shown that the expectation value in the false
vacuum is decreasing exponentially in time. In the next step, we
are interested in the expectation value in the non-perturbative
thermal quantum state. According to the authors of
\cite{starobinsky,berrerarev,int,habathermal}, in a slow roll
approximation, the field evolution in such a quantum state can be
obtained as a solution of a stochastic equation with quantum and
thermal noise. We discuss approximate solutions of such equations.
We conclude that the approximate solutions have the same
qualitative (exponential) behavior as the classical solutions. We
are unable at the moment to treat precisely the non-linear
stochastic equations or the Fokker--Planck partial differential
equations for the probability distribution of the stochastic
process. The large time behavior is determined by the equation for
stationary probability distribution. We can obtain an exact
solution of this equation. This allows   calculating  the
expectation values at large time. The classical slow-roll
approximation is not reliable for large time. We   showed  in
\cite{hepj} that the stochastic version  (because of quantum and
thermal fluctuations) can make sense for arbitrarily large time.
Then, the evaluation of the mean value of the energy density may
give some qualitative explanation why the cosmological term is big
at the early time of inflation and it is decreasing to a small
value at large time. The same model when applied in the $\delta N$
formalism (initiated in \cite{starobmulti,flu2} and developed in
\cite{ven}) can be used for a calculation of the spectrum of
quantum and thermal fluctuations.

 The       paper is  organized as  follows. In Section \ref{sec.2}, we define the
 semi-classical approximation for Einstein equations. In Section \ref{sec.3},
 we discuss the slow-roll stochastic approximation for quantum
evolution in a thermal state. In Section \ref{sec.4}, we calculate
the energy-density in the absence of noise. In Section
\ref{sec.5}, the approximate stochastic equations are solved. The
asymptotic behavior of expectation values is discussed in Section
\ref{sec.6} where the stationary probability for the stochastic
process is calculated. In particular, we show that the
cosmological constant at large time is proportional to the thermal
diffusion constant.

\section{The Semi-Classical Einstein Equations} \label{sec.2}
First, we quantize the scalar field in an external metric.
Subsequently, we determine the metric from the Einstein equations
\begin{equation}
R_{ \mu\nu}-\frac{1}{2}g_{\mu\nu}R=8\pi G<\psi\vert
T_{\mu\nu}(\phi_{t})\vert\psi>.
\end{equation}

Here, $G$ is the Newton constant, $\phi_{t}$ is the quantum scalar
field in an external metric $g_{\mu\nu}$, and $\vert\psi>$ is the
quantum state of the field $\phi$. It is understood that Equation
(1) comes from an average of a complete inflationary model like
that of Ref. \cite{brand}.
 The
quantization of renormalizable interactions in an external metric
is well-defined on the perturbative level \cite{birrel}. The
expectation values of the energy-momentum are usually calculated
in the Bunch--Davis vacuum of the free field theory \cite{birrel}.
They require renormalization because of the ultraviolet
divergences. The expectation value and renormalization add new
terms to the lhs of the Einstein equations (Equation  (1))
\cite{s1,s2}. In this paper, we separate the long wave and short
wave problems, restricting ourselves to the study of the latter.
  To discuss the ground state of an interacting field
theory, we need to go beyond the perturbation theory.

 We would also be
  interested in an expectation value of the energy-momentum
tensor $T_{\mu\nu}$ on the rhs of Equation (1) in a mixed state
$\rho$ , i.e.,
\begin{equation}
R_{\mu\nu}-\frac{1}{2}g_{\mu\nu}R=8\pi G Tr\Big(\rho
T_{\mu\nu}(\phi_{t})\Big).
\end{equation}

In the slow-roll classical approximation, the expectation value of
Equation (1) is
\begin{equation}
<\psi\vert T_{\mu\nu}(\phi_{t}(\phi))\vert\psi>=\int d\phi \vert
\psi(\phi)\vert^{2}T_{\mu\nu}(\phi_{t}(\phi)),
\end{equation}
where $\phi_{t}(\phi)$ is the solution of the equation
\begin{equation}
3H\partial_{t}\phi= -V^{\prime}(\phi).
\end{equation}
with the initial condition $\phi$. It follows from the Einstein
equations (Equation  (1), in the Friedman form) that in a flat
homogeneous expanding universe
\begin{equation}
H^{2}=\frac{8\pi G}{3}V(\phi)
\end{equation}
in the slow-roll approximation.

\section{Stochastic Equations for Slow-Roll
Inflation} \label{sec.3} Observations of the CMB radiation show
that the universe is spatially flat. We consider a flat
(homogeneous) expanding metric
\begin{displaymath}
ds^{2}=g_{\mu\nu}dx^{\mu}dx^{\nu}=dt^{2}-a(t)^{2}d{\bf x}^{2}.
\end{displaymath}

In this metric, the quantum field behaves as a classical diffusion
process \cite{starobinsky,vilenkin} with the noise
$\frac{3}{2\pi}H^{\frac{5}{2}}\partial_{t}W$ \cite{starobquan}
where $W $ is the Brownian motion \cite{ikeda}. We consider a
stochastic wave equation with friction
\begin{equation}
\partial_{t}^{2}\phi-a^{-2}\triangle\phi+(3H+\beta
\gamma^{2})\partial_{t}\phi+V^{\prime}(\phi)+\frac{3}{2}\beta\gamma^{2}H\phi=\eta,
 \end{equation}
where $\beta^{-1}$ is the temperature of the environment,
$H=a^{-1}\partial_{t}a$ and $\eta$ is a noise. Further on, we make
the slow-roll approximation neglecting $\partial_{t}^{2}\phi$
term. We assume that the friction $\beta\gamma^{2}$ is small in
comparison to $3H$. We also neglect the spatial derivatives
($\triangle\phi$).
 We can take into account thermal effects by an addition of the
thermal noise to the rhs of Equation (1). Now,
\begin{equation}
\eta=\gamma
a^{-\frac{3}{2}}\partial_{t}B+\frac{3}{2\pi}H^{\frac{5}{2}}\partial_{t}W,
\end{equation}
where the first term describes the thermal noise and the second
term the quantum noise. The thermal part of this equation is
derived   in \cite{berrera,habaadv}. The factor $a^{-\frac{3}{2}}$
comes from $\det\vert g_{\mu\nu}\vert ^{\frac{1}{4}}$ and the
factor $H^{\frac{5}{2}}$ in the Starobinsky noise is chosen     to
reproduce the correlation functions of the quantum scalar field in
an expanding universe \cite{starobquan}. The (white) noise
$\partial_{t}{B}$ is the Gaussian random process with the
covariance
\begin{equation}
\langle
\partial_{t}B\partial_{s}B\rangle=\delta(t-s).\end{equation}
$\partial_{t}W$ is an independent Gaussian stochastic process with
the same covariance (Equation (8)). Equation (1) with quantum and
thermal noise is
  discussed in \cite{ramos,int,habathermal}.

We interpret Equation (6) as a stochastic equation in the
Stratonovitch sense \cite{ikeda} (we apply the stochastic calculus
and the $\circ$-circle notation of   \cite{ikeda} for stochastic
multiplication). In the slow-roll approximation and small friction
($\gamma^{2}\simeq 0$),

\begin{equation}
3Hd\phi = -V^{\prime}dt +\gamma a^{-\frac{3}{2}}\circ
dB+\frac{3}{2\pi}H^{\frac{5}{2}}\circ dW.
\end{equation}

If $H$ is known (from Equation (5)), then in principle we can
determine $a(\phi)$ after solving the stochastic Equations (1) and
(6). We can obtain an explicit formula if we neglect the noise and
apply the slow roll approximation in Equation (9) (without noise).
Then,
\begin{equation}
\ln (a)=\int Hdt=\int d\phi (\frac{d\phi}{dt})^{-1}H =-8\pi G
 \int d\phi V(V^{\prime})^{-1}.
 \end{equation}

We can generalize the stochastic Equation  (9) to multiple scalar
fields \cite{starobmulti,venn3} $\phi=(\phi^{1}, ...,\phi^{D})$ as
in the standard model of weak and strong interactions with the
Higgs potential

\begin{displaymath}
V=\frac{g}{4}(\vert\phi\vert^{2}-\frac{\mu^{2}}{g})^{2}.
\end{displaymath}

 Then, the noise $\eta=(\eta^{1},....,\eta^{D})$
 consists of independent random Gaussian variables with
the same variance. In Equation (9) $\phi$ and the noises should be
treated as vectors, $V^{\prime}\rightarrow \nabla V$ and
$(\partial_{t}\phi)^{2}\rightarrow \vert
\partial_{t}\phi\vert^{2}$ (the length of the vector $\partial_{t}\phi$). We can calculate $a(\phi)$ if
$V=V(\vert\phi\vert)$ is rotation invariant with
$\vert\phi\vert^{2}=(\phi^{1})^{2}+...+(\phi^{D})^{2}$.
 Then   (after an omission of noise in Equation (9)),
\begin{equation}\ln (a)=\int Hdt=\int d\vert\phi\vert
(\frac{d\vert\phi\vert}{dt})^{-1}H =-8\pi G
 \int d\vert\phi\vert V\Big(\frac{dV}{d\vert\phi\vert}\Big)^{-1}.
 \end{equation}

 We have the Fokker--Planck equation \cite{risken} for the
probability distribution of $\phi$  (with the Stratonovich
interpretation of the stochastic equations \cite{ikeda})
\begin{equation}\begin{array}{l}
\partial_{t}P=\frac{\gamma^{2}}{18}\partial_{\phi}\frac{1}{Ha^{\frac{3}{2}}}\partial_{\phi}\frac{1}{Ha^{\frac{3}{2}}}P
+\frac{1}{8\pi^{2}}\partial_{\phi}H^{\frac{3}{2}}\partial_{\phi}H^{\frac{3}{2}}P
+\partial_{\phi}(3H)^{-1}V^{\prime}P.  \end{array}\end{equation}

In the multifield case ($\phi\in R^{D}$) with
$\partial_{j}=\frac{\partial}{\partial\phi^{j}}$, Equation (12)
reads
\begin{equation}\begin{array}{l}
\partial_{t}P=\sum_{j}\partial_{j}\Big(\frac{\gamma^{2}}{18}\frac{1}{Ha^{\frac{3}{2}}}\partial_{j}\frac{1}{Ha^{\frac{3}{2}}}P
+\frac{1}{8\pi^{2}}H^{\frac{3}{2}}\partial_{j}H^{\frac{3}{2}}P
+(3H)^{-1}\partial_{j}V\Big) P.  \end{array}\end{equation}

We express $H(\phi)$ as a function of $\phi$ from Equation (5).
The dependence of $a$ on $\phi$ is more involved. We determine it
in the slow-roll approximation in Equation (10). All the formulas
in this section can be generalized to a multifield case (Equation
(11)) just by a replacement $\phi\rightarrow\vert\phi\vert$.
 For the double-well potential (treated as an approximate
 realization of the models of inflation \cite{bard,ring}, e.g., as
 an approximation to the Starobinsky model \cite{trot,ketov}),
\begin{equation}
V(\phi)=\frac{g}{4}(\phi^{2}-\frac{\mu^{2}}{g})^{2} ,
\end{equation}
\begin{equation}\begin{array}{l}
a=\vert\phi\vert^{\frac{2\pi G\mu^{2} }{g}}\exp(-\pi G\phi^{2}).
\end{array}\end{equation}

If $a\rightarrow 0$, then either $\phi\rightarrow 0$ or
$\phi\rightarrow\infty$.
 According to    Equation  (9) ($B=W=0$),
the classical slow roll time evolution of $\phi$ is determined by
\begin{equation}
-\sqrt{24\pi
G}\Big(\frac{g}{4}(\phi^{2}-\frac{\mu^{2}}{g})^{2}\Big)^{\frac{1}{2}}\frac{d\phi}{dt}=g\phi(\phi^{2}-\frac{\mu^{2}}{g}).
\end{equation}

Hence, if $\mu g^{-\frac{1}{2}}\geq \phi\geq 0$, then $\phi$ is
increasing. If $\phi\geq \mu g^{-\frac{1}{2}}$, then $\phi$ is
decreasing to $\mu g^{-\frac{1}{2}}$. It follows that
$\phi(t)\rightarrow \mu g^{-\frac{1}{2}}$. In reality, the ground
state is beyond the slow roll regime as the slow-roll parameters
\begin{displaymath}
\tilde{\epsilon}=\frac{1}{16\pi G}(\frac{V^{\prime}}{V})^{2}=
\frac{1}{16\pi G}\phi^{2}(\phi^{2}-\frac{\mu^{2}}{g})^{-2}
\end{displaymath}
and \begin{displaymath}\tilde{\eta}=\frac{1}{8\pi
G}\frac{V^{\prime\prime}} {V}= \frac{1}{2\pi
G}(3\phi^{2}-\frac{\mu^{2}}{g})(\phi^{2}-\frac{\mu^{2}}{g})^{-2}.
\end{displaymath}
tend to infinity close to the ground state. Nevertheless, the
stochastic dynamical system in Equation (9) still makes sense
close to the minimum of the potential but presumably does not
approximate the second-order wave equation (6). As     shown in
Section \ref{sec.5} for double-well potentials (for a general
discussion, see \cite{int}), the stochastic equation with the
Starobinsky noise leads to a non-integrable stationary probability
distribution (if treated as a system on the whole real line),
whereas the system with the thermal noise has a normalizable
stationary distribution. The fact that the solution of the
equation for the stationary probability distribution is not
normalizable could be ignored. It appears in a range of $\phi$
where the restrictions on the parameters $\tilde{\epsilon}$ and
$\tilde{\eta}$ are violated. For deterministic systems, we can
restrict the initial values of the field and the time evolution in
order to satisfy the requirement of small $\tilde{\epsilon}$ and
$\tilde{\eta}$. It is more difficult to do it in a random system
because the noise can move the system to the forbidden region of
the field configurations. We can define the stochastic system in a
required domain of field configurations by an imposition of
boundary conditions (as discussed in \cite{venn3,venn} for the
Starobinsky stochastic equation with $\gamma=0$). However, in such
a case, the stochastic process and its probability distribution
depend on the boundary conditions. Nevertheless, some correlation
functions may have a negligible dependence on boundary conditions,
as discussed in \cite{venn3,venn} (for more on boundaries in
diffusions, see \cite{ikeda}). As a result of fluctuations, the
stochastic process does not feel the singularity
$g\phi^{2}=\mu^{2}$. The requirement of small values of
$\tilde{\epsilon}$ and $\tilde{\eta}$ can be treated as the
requirements of a strong friction which is violated when $H\simeq
\vert \phi^{2}-\frac{\mu^{2}}{g}\vert$ in Equation (6). The
quantum noise in Equation (7) (which is multiplied by $H$) is also
vanishing at $H(\phi)=0$. It is known in the theory of Brownian
motion \cite{nelson} that under the assumption of a strong
friction the random second-order differential Equation (6) can be
replaced by the first-order Equation (9). It may remain true with
the thermal noise in Equation (7) which is not vanishing. In any
case, the presence of the thermal noise allows an asymptotic
regular behavior of the solutions of $\phi_{t}$, as is exhibited
by the existence of the stationary probability (proved in Section
\ref{sec.6}). Hence, we expect that the stochastic slow-roll
process in Equation  (9) may well approximate the second-order one
despite   the increase of the small-roll parameters
$\tilde{\epsilon}$ and $\tilde{\eta}$ at $g\phi^{2}=\mu^{2}$.

 Equation (16) is non-analytic at $g\phi^{2}=\mu^{2}$. As an
example of a potential which does not lead to a singular slow-roll
equation, we consider
\begin{equation}
V(\phi)=\frac{g_{4}}{4}(\phi^{2}-\frac{\mu^{2}}{g})^{4} ,
\end{equation} where $g_{4}$ is a dimensional constant. Then,
\begin{equation}\begin{array}{l}
a(\phi)=\vert\phi\vert^{\frac{\pi
G\mu^{2}}{g}}\exp(-\frac{1}{2}\pi G\phi^{2}).
\end{array}\end{equation}

\section{Expectation Value of the Energy-Momentum in the Semi-Classical Approximation}\label{sec.4}

In Equation (1), it is understood that the evolution equation for
the quantum field $\phi_{t}$ has been solved. Such an equation can
be formulated by imposing the canonical commutation relations at
the initial time. In static coordinates, it may be possible to
define a Hamiltonian which generates the time evolution (and
allows defining  a thermal state). If there is a Hamiltonian
${\cal H}$ generating a unitary time evolution $\exp(-i{\cal
H}t)$(at least at small time), then, decomposing the expectation
value in Equation (1) in a complete set of eigenfunctions $\vert
E>$ of ${\cal H}$, we get
\begin{equation}\begin{array}{l}<\psi\vert
V(\phi_{t}(\phi))\vert\psi>= <\psi_{t}\vert
V(\phi)\vert\psi_{t}>\cr=\int_{0}^{\infty}dE\int_{0}^{\infty}dE^{\prime}
<\psi\vert E><E\vert V(\phi)\vert E^{\prime}>
<E^{\prime}\vert\psi>\exp(-i(E^{\prime}-E)t)\cr=\int_{-\infty}^{\infty}d\epsilon
\gamma(\epsilon)\exp(-i\epsilon t),
\end{array}\end{equation}
where we introduce  \begin{displaymath}\epsilon=E^{\prime}-E
\end{displaymath}
and
\begin{equation}
\gamma(\epsilon)=\int_{0}^{\infty}dE
<E+\epsilon\vert\psi><\psi\vert E><E\vert V(\phi)\vert
E+\epsilon>.
\end{equation}

Here, $\vert E>$ is the eigenstate of the Hamiltonian with an
eigenvalue $E\geq 0$. The behavior of the Fourier transform
depends on $\gamma(\epsilon)$. Note, however, that $\epsilon $ is
unbounded from below and the Paley--Wiener theorem is not
applicable. We could also write    Equation  (19) in the form
\begin{equation}\begin{array}{l}
<\psi\vert
V\Big(\phi_{t}(\phi)\Big)\vert\psi>=\sum_{n,m}<\psi_{t}\vert\chi_{n}><\chi_{m}\vert\psi_{t}><\chi_{n}\vert
 V(\phi)\vert \chi_{m}>,
\end{array}\end{equation}
where $\chi_{n}$ is a basis of eigenstates. Then, the
Paley--Wiener theorem can be applied estimating each term  (as in
\cite{krauss,ssu}) in the series in Equation (21) but the behavior
of the infinite sum remains unclear. The integrals in Equation
(20) are calculable in quantum mechanics if we solve the
eigenvalue problem in a given potential $V$. However, it is rather
difficult to determine the time behavior of the Fourier integrals
in general.
   Equation  (19) may be problematic in the case of a time
dependent metric. Nevertheless, we can see that the exponential
decay in Equation (19) (which we       prove below) is not
forbidden by the Paley--Wiener theorem even in an ideal case of
the Minkowski metric.

We calculate the expectation value of the energy-momentum in some
approximations. We hope that it correctly describes the relevant
regime of the vacuum decay. We consider the potential in Equation
(14). Then, Equation (16) can be expressed as
\begin{equation}
-\sqrt{6\pi Gg}\frac{d\phi}{dt}=g\phi \epsilon
(g\phi^{2}-\mu^{2}),
\end{equation} where $\epsilon$ is an antisymmetric function with $\epsilon(y)=1$
 for $y>0$ and $\epsilon(0)=0$. Equation (22) is discontinuous at the minimum
 of the potential.Let
\begin{equation}
Y(\phi)=\phi^{2}-\frac{\mu^{2}}{g}
\end{equation} and
\begin{displaymath}
 \alpha=\sqrt{\frac{g}{6\pi G}}.\end{displaymath}

 Then, Equation (22) reads
\begin{displaymath}
\partial_{t}Y=-2\alpha
Y\epsilon(Y)-\frac{2\alpha\mu^{2}}{g}\epsilon(Y)
\end{displaymath}
with the solution $\phi_{t}(\phi)$ for $g\phi^{2}>\mu^{2}$
\begin{equation}
\phi_{t}(\phi)^{2}=\exp(-2\alpha t)\phi^{2}
\end{equation} as long as $\exp(-2\alpha t)\phi^{2}\geq \frac{\mu^{2}}{g}$
and the solution $\phi_{t}(\phi)$ for $g\phi^{2}<\mu^{2}$
\begin{equation}
\phi_{t}(\phi)^{2}=\exp(2\alpha t)\phi^{2}
\end{equation} as long as $\exp(2\alpha t)\phi^{2}\leq \frac{\mu^{2}}{g}$.
Let us note that a solution exists only for a limited time
interval $[0,t]$ (depending on $\phi$)
\begin{displaymath}
\exp(2\alpha t)\leq \frac{\mu^{2}}{g\phi^{2}}
\end{displaymath}
if $g\phi^{2}<\mu^{2}$ and
\begin{displaymath}
\exp(-2\alpha t)\geq \frac{\mu^{2}}{g\phi^{2}}
\end{displaymath}if $g\phi^{2}>\mu^{2}$.

 The solutions in Equations (24) and (25) exist until  the time when $\phi_{t}$
achieves the minimum. However, when $g\phi^{2}$ is close to
$\mu^{2}$,   the slow roll conditions of small $\tilde{\epsilon}$
and small $\tilde{\eta}$ are violated. Hence, close to the minimum
of the potential, the solution $\phi_{t}(\phi)$ does not
approximate the solutions of the wave in Equation (6). In fact,
the solutions $\phi_{t}(\phi)$ of Equation (6) oscillate in the
vicinity of the minimum of the potential, whereas Equations (24)
and  (25) decay to the minimum. We can check whether the second
time derivative in Equation (6) is negligible, i.e., if $\vert
\partial_{t}^{2}\phi\vert<<3H\vert\partial_{t}\phi\vert$.   From Equations (24) and (25), we
obtain that this is the case if $\alpha<<3H$, i.e., the decay of
$\phi_{t}$ should be slow in comparison to the Hubble expansion.
From Equation (5), we obtain
\begin{equation}
\vert \phi^{2}-\frac{\mu^{2}}{g}\vert >>\frac{1}{6\pi
G}=\frac{4}{3}m_{PL}^{2}.
\end{equation}

Comparing with the definitions of $\tilde{\epsilon} $ and
$\tilde{\eta}$, we can see that $\tilde{\epsilon} $ and
$\tilde{\eta}$ will be small if $\phi^{2}<<m_{PL}^{2}$ and if the
condition in Equation (26) is satisfied or if
$\frac{\mu^{2}}{g}>>m_{PL}^{2}$.

We     calculate a decay of the expectation value of the
energy-momentum in a state $\psi$ which is not a ground state. Let
us consider a wave function concentrated at $\phi=0$
\begin{equation}
\psi=\Big(\frac{\sigma}{2\pi}\Big)^{\frac{1}{4}}\exp(-\frac{\sigma\phi^{2}}{4})
\end{equation} with a certain
$\sigma>0$. The expectation value in Equation (19) is
\begin{equation}\begin{array}{l}
<\psi\vert V\Big(\phi_{t}(\phi)\Big)\vert \psi>=
\Big(\frac{2\pi}{\sigma}\Big)^{\frac{1}{2}}\frac{g}{4}
\Big(\int_{\sqrt{\frac{1}{g}}\exp(\alpha
t)\mu}^{\infty}d\phi\exp(-\frac{\sigma \phi^{2}}{2})(\exp(-2\alpha
t)\phi^{2}-\frac{\mu^{2}}{g})^{2}\cr +
\int_{0}^{\sqrt{\frac{1}{g}}\exp(-\alpha
t)\mu}d\phi\exp(-\frac{\sigma \phi^{2}}{2})(\exp(2\alpha
t)\phi^{2}-\frac{\mu^{2}}{g})^{2}\Big),
\end{array}\end{equation}where as $\phi_{t}(\phi)$ we insert  the
solutions in Equations  (24) and (25). We can express the first
integral in Equation (28) by the probability function $\Phi$
\cite{grad}\begin{equation}
\int_{v}^{\infty}du\exp(-\frac{u^{2}}{4\beta})=\sqrt{\pi\beta}
\Big(1-\Phi(\frac{v}{2\sqrt{\beta}})\Big).
\end{equation}

It decays  more quickly  than exponentially. The second integral
can be estimated by an elementary change of variables. We obtain
\begin{equation} <\psi\vert V(\phi_{t}(\phi))\vert \psi>=<\psi\vert
V(\phi)\vert \psi>\exp(-\alpha t)(1+K(t))
\end{equation}(with a bounded function $K(t),K(0)=0$) showing the exponential decay of the cosmological term.

Next, let us consider a wave function $\chi$ concentrated around
the ground state at $\vert\phi\vert=\frac{\mu}{\sqrt{g}}$ such
that
\begin{equation}
\chi=const
\end{equation} in the interval
\begin{equation} \frac{\mu}{\sqrt{g}}-r\leq
\vert\phi\vert\leq\frac{\mu}{\sqrt{g}}+r
\end{equation} with a certain $r\simeq m_{PL}$ (according to Equation (26)) and $\chi=0$ outside this interval.
Then, for sufficiently small r (depending on $t$), the only
solution of the classical equation is the constant solution with
the initial condition $\vert\phi\vert=\frac{\mu}{\sqrt{g}}$. The
contribution of this solution to the expectation value is
\begin{equation}\begin{array}{l}
<\chi_{t}\vert V(\phi)\vert \chi_{t}>=0.
\end{array}\end{equation}

We can conclude that the expectation value of the energy in a
state concentrated at the false vacuum of $\phi$ is decaying in
time, whereas the expectation value concentrated at the ground
state value of $\phi^{2}$ (at $\frac{\mu^{2}}{g}$) is zero.

 It is
not straightforward to compare the result in Equation (30) with
other studies of the quantum scalar field in expanding universes.
This has been usually done in de Sitter space with $H=const$,
whereas, in our Einstein--Klein--Gordon system, $H$ is determined
by the field $\phi_{t}$ in Equation (5). Note that if (according
to Equation (24)) $\phi_{t}^{2}=\exp(-2\alpha t)\phi^{2}$ for
$g\phi^{2}>\mu^{2}$, then $H$ is decreasing exponentially to zero.
The same is true for $g\phi^{2}<\mu^{2}$ when
$\phi_{t}^{2}=\exp(2\alpha t)\phi^{2}$ (according to Equation
(25)). We can compare our results with the ones concerning quantum
field theory with $H=const$. Thus, the decay rate $\alpha$ of
Equations (24) and (30) coincides with the one of Starobinsky and
Yokoyama (\cite{starobquan}, Equation (60)) if in Equation (5) for
$H$ we insert $\phi=0$ (the local maximum) of the potential.

Equation (16) being piece-wise linear may look rather special. For
the potential in Equation (17), we obtain the slow-roll equation
\begin{equation}
\frac{d\phi}{dt}=-2\alpha_{4}\phi(\phi^{2}-\frac{\mu^{2}}{g}),
\end{equation}where
\begin{displaymath}
 \alpha_{4}=\sqrt{\frac{g_{4}}{6\pi G}}.\end{displaymath}

Its solution with the initial condition $\phi$ is
\begin{equation}
\phi_{t}(\phi)^{2}=\frac{\mu^{2}}{g}\phi^{2} \Big(
\phi^{2}(1-\exp(-\frac{4\alpha_{4}\mu^{2}t}{g}))+\frac{\mu^{2}}{g}\exp(-\frac{4\alpha_{4}\mu^{2}t}{g})\Big)^{-1}.
\end{equation}

When we insert  Equation (35) into the expectation value in
Equation (19) with the Gaussian wave function in Equation (27)
concentrated at $\phi=0$,
  there will be an unphysical contribution from an unstable
fixed point $\phi_{t}$ with the initial condition $\phi\simeq 0$.
  To eliminate this contribution, we assume that
$\psi(\phi)=0$ for $\vert \phi\vert\leq r$; then, we obtain an
estimate\begin{equation} <\psi\vert V(\phi_{t})\vert \psi>\leq
K(r)\exp(-\frac{16\alpha_{4}\mu^{2}t}{g}).
\end{equation}

 If $\psi$ is sharply concentrated at $g\phi^{2}=\mu^{2}$ with the variance $\frac{1}{\sigma}\rightarrow 0$,
then

$<\psi_{t}\vert V(\phi)\vert \psi_{t}>\simeq
\frac{1}{\sigma^{4}}\rightarrow 0$. In the next sections, we
calculate the expectation values resulting from quantum and
thermal fluctuations at large time.
\section{Expectation Value of the Energy-Momentum with the
Stochastic Slow Roll Approximation} \label{sec.5} We neglect the
second-order time derivatives but we take into account the quantum
and thermal noise. Now, Equation (9) for "quantum" $ \phi$ in a
thermal state reads

\begin{equation}
\sqrt{6\pi g G}\frac{d\phi}{dt}=-g\phi \epsilon
(g\phi^{2}-\mu^{2})+\gamma a^{-\frac{3}{2}}\vert
\phi^{2}-\frac{\mu^{2}}{g}\vert^{-1}\frac{dB}{dt}+\frac{3}{2\pi}\vert
\phi^{2}-\frac{\mu^{2}}{g}\vert^{\frac{3}{2}}(\frac{8\pi
Gg}{3})^{\frac{5}{4}}\frac{dW}{dt}.
\end{equation}

The approximations of this equation depend on whether $\phi$ is
close to the minimum of $V$ or not. For a small $\phi$
($\phi^{2}<<\frac{\mu^{2}}{g}$), we have
\begin{equation}
d\phi=\alpha\phi dt
+\frac{\alpha\gamma}{\mu^{2}}a^{-\frac{3}{2}}\circ
dB+\frac{3\alpha\mu^{3}}{2\pi }(\frac{8\pi
G}{3g})^{\frac{5}{4}}\circ dW.
\end{equation}

From    Equation   (15), $a^{-\frac{3}{2} }\simeq \phi
^{-\frac{3\pi G\mu^{2}}{g}}$.   Hence, this term is large for a
small $\phi$, whereas the last term in Equation (38) is negligible
in comparison to the thermal one. Let us consider $Q=\phi^{m}$.
Then,
\begin{equation}
dQ=m\phi^{m-1}\circ d\phi=\alpha m Q dt
+\frac{\alpha\gamma}{\mu^{2}}mQ^{\frac{m-1-\frac{3\pi
G\mu^{2}}{g}}{m}}\circ dB.
\end{equation}

If we choose \begin{equation}m=1 +\frac{3\pi G\mu^{2}}{g},
\end{equation}
then $Q$ is the Ornstein--Uhlenbeck process \cite{ikeda}
\begin{equation}
Q_{t}=\exp(m\alpha
t)\phi^{m}+\frac{\alpha\gamma}{\mu^{2}}m\int_{0}^{t}\exp(m\alpha(t-s))dB_{s}.
\end{equation}

Equation (38) allows   calculating $\vert\phi\vert =\vert
Q\vert^{\frac{1}{m}}$. From the correlation functions of the
Ornstein--Uhlenbeck process, we can conclude that the classical
behavior $\phi\simeq \exp(\alpha t)$ is strengthened by the noise.
This can already be seen when calculating the covariance of
$Q_{t}$
\begin{equation}
\langle Q_{t}^{2}\rangle =\exp(2m\alpha
t)\phi^{2m}+\frac{m\gamma^{2}}{2\mu^{4}\alpha}(\exp(2m\alpha
t)-1).
\end{equation}

 We can also calculate the expectation value in a thermal state
$\rho$ using the formula \begin{equation} Tr\Big(\rho
V\Big(\phi_{t}(\phi)\Big)\Big)=\int d\phi d\phi^{\prime}
P_{0}(\phi)P_{t}(\phi,\phi^{\prime})V(\phi^{\prime}),\end{equation}
where $P_{t}$ is the transition function for the stochastic
process $\phi_{t}$ (Equation (38)) (see \cite{habaadv} for
Equation  (43)). $P_{0}$ is the distribution of the initial value.
For a thermal state $\rho$ and the Ornstein--Uhlenbeck $Q$, we
would have
\begin{displaymath}
P_{0}(Q)=\exp(-\frac{m\alpha}{\hbar} Q\tanh(\frac{\hbar\beta
m}{2}\alpha)Q),
\end{displaymath}
where $\beta$ is the inverse temperature. We would have to express
$\phi_{t}$ by $Q_{t}$ in Equation (43) in order to obtain the
transition function for $\phi_{t}$ in terms of the known
transition function for the Ornstein--Uhlenbeck process $Q_{t}$.
With the result in Equation (42), the calculations of  Equation
(43) confirm the growth of $\vert\phi\vert$ from its small initial
value.

If $g\phi^{2}>>\mu^{2}$, then there is a damping negative power of
$\phi$ in $a^{-\frac{3}{2}}$ in the thermal term (which is
dominating at large $\frac{\mu^{2}}{g}$ before the exponential in
Equation (15) becomes significant). When we neglect the thermal
term in Equation (37),
  the stochastic equation reads
\begin{equation}
d\phi=-\alpha\phi dt+\lambda\phi^{3}\circ dW,
\end{equation}
where
\begin{displaymath}
\lambda=\frac{3\alpha}{2\pi g}(\frac{8\pi Gg}{3})^{\frac{5}{4}}.
\end{displaymath}

Let
\begin{displaymath}
\Omega=\phi^{-2}.
\end{displaymath}

Then,
\begin{equation}
d\Omega=2\alpha \Omega dt-2\lambda dW.
\end{equation}

Hence,
\begin{equation}
\Omega_{t}= \phi^{-2}\exp(2\alpha
t)-2\lambda\int_{0}^{t}\exp(2\alpha(t-s)) dW_{s}.
\end{equation}

The expectation value of the potential could again be determined
by the rhs of    Equation  (43) by expressing $\phi_{t}$ by
$\Omega_{t}$ and calculating the expectation values of
$\Omega_{t}$ by means of the known transition function for the
Ornstein--Uhlenbeck process (as a result, $\phi_{t}\simeq
\exp(-\alpha t)$). In the limit $\lambda\rightarrow 0$, the
transition function $P_{t} $ tends to the $\delta$ function. Then,
with $P_{0}=\vert\psi\vert^{2}$, we return to    Equation  (30)
for the expectation value of the previous section.
 In another way, we can see that the
quantum noise supports the classical behavior derived in Section
\ref{sec.5} as from Equation (46) it follows (we can derive such
expectation values for any power of $\Omega_{t}$)
\begin{equation}
\langle \Omega_{t}^{2}\rangle=\phi^{-4}\exp(4\alpha t)+
\frac{\lambda^{2}}{\alpha}(\exp(4\alpha t)-1).
\end{equation}

According to Equation (47),$\phi_{t}^{-2}$ is growing in time;
hence, $\langle V\rangle$ is decreasing in time. As     shown in
the next section, it is decreasing to the expectation value in a
stationary state. Equations (42) and (47) show that the
corrections resulting from thermal as well as quantum fluctuations
lead to the same decay law $\exp(-2\alpha t)$  (appearing also in
\cite{starobquan}, Section IVC). Only the amplitude of $\phi$
fluctuation is adding to the classical $\phi^{2}$.

 The stochastic equation for the potential in Equation (17) takes the
form
\begin{equation}\begin{array}{l}
d\phi=-2\alpha_{4}\phi(\phi^{2}-\frac{\mu^{2}}{g})+
\frac{\alpha_{4}\gamma}{g}a^{-\frac{3}{2}}(\phi^{2}-\frac{\mu^{2}}{g})^{-2}\circ
dB\cr +\frac{3\alpha_{4}}{2\pi g}\Big(\frac{2\pi
gG}{3}\Big)^{\frac{5}{4}}\vert\phi^{2}-\frac{\mu^{2}}{g}\vert^{3}\circ
dW.
\end{array}\end{equation}

For small $\phi$, the thermal noise proportional to
$a^{-\frac{3}{2}}$ dominates over the quantum noise. We can write
an approximation of Equation (48) using    Equation  (18) for
$a^{-\frac{3}{2}}$ (neglecting the exponential for a small $\phi$)
as
\begin{equation}
d\phi =\frac{2\alpha_{4}\mu^{2}}{g}\phi dt +\frac{\alpha_{4}\gamma
g}{\mu^{4}}\vert\phi\vert^{-\frac{3\pi G\mu^{2}}{2g}}\circ dB.
\end{equation}

Let \begin{equation} q=\vert\phi\vert^{r}
\end{equation}
with \begin{equation} r=1+\frac{3\pi G\mu^{2}}{2g}
\end{equation}

Then,
\begin{equation}
dq=\frac{2\alpha_{4} r\mu^{2}}{g}qdt+\frac{r\alpha_{4}\gamma
g}{\mu^{4}}dB.
\end{equation}

The solution of Equation (52) is the Ornstein--Uhlenbeck process.
We can calculate all correlation functions as in Equation (42) and
conclude that, for a small $\phi$, we have
$\phi_{t}(\phi)^{2}\simeq \exp(\frac{4\alpha_{4}\mu^{2}}{g}t)$.
This conclusion is in agreement with the classical solution in
Equation (35) when we expand it in $\phi^{2}$.

\section{Fokker--Planck Equation and Its Stationary Probability Distribution }\label{sec.6}

 Environmental noise is present in
all physical systems. Its crucial role in equilibration of
dynamical systems is well-known \cite{ruell}. Its action can be
seen as a stabilization. Some correlation functions have a
singular or chaotic behavior at $t\rightarrow \infty$ without
noise and a smooth behavior with noise. We could approach the
calculation of the expectation value of the energy-momentum by
means of the Fokker--Planck equation for the transition
probability (12). When the initial state is far from the ground
state,   we have   the exponential decay of expectation values in
the semi-classical approximation in Section \ref{sec.4}. As a
result of thermal and quantum fluctuations, there will be a
non-zero correction to this result. It is rather difficult to
obtain an exact solution of the Fokker--Planck equation (Equation
(12)) or even to estimate a qualitative behavior of its solutions.
However, the large time limit can be obtained from the equation
for the stationary distribution.

 The stationary probability $P_{\infty}(\phi)$ is the limit of $P_{t}$ for $t\rightarrow\infty$ \cite{risken}
 (it does not depend on the initial condition). It can be obtained from the
 requirement $\partial_{t}P=0$, which gives (with
 the Stratonovitch interpretation and after an integration
 over $\phi$)\begin{equation}\begin{array}{l}
\frac{\gamma^{2}}{18}\frac{1}{Ha^{\frac{3}{2}}}\partial_{\phi}\frac{1}{Ha^{\frac{3}{2}}}P
+\frac{1}{8\pi^{2}}H^{\frac{3}{2}}\partial_{\phi}H^{\frac{3}{2}}P
+(3H)^{-1}V^{\prime}P =0.\end{array}\end{equation}

 In the
multidimensional case in Equation (13), the requirement
$\partial_{t}P=0$ is satisfied (for functions $P$ and $V$
depending only on $\vert \phi\vert$) if
\begin{equation}\begin{array}{l}
\frac{\gamma^{2}}{18}\frac{1}{Ha^{\frac{3}{2}}}\partial_{\vert\phi\vert}\frac{1}{Ha^{\frac{3}{2}}}P
+\frac{1}{8\pi^{2}}H^{\frac{3}{2}}\partial_{\vert\phi\vert}H^{\frac{3}{2}}P
+(3H)^{-1}P \partial_{\vert\phi\vert}V=0.\end{array}\end{equation}

Let us consider the simplest case first. The stationary solution
of Equation (53) without the Starobinsky (quantum) noise is
\begin{equation}\begin{array}{l}
 P_{\infty}\equiv \sqrt{V}a^{\frac{3}{2}}\exp(-\gamma^{-2}F(\phi))=\sqrt{V}\exp\Big(-12\pi
G\int^{\phi}d\phi^{\prime}
(V^{\prime})^{-1}V\Big)\cr\times\exp\Big(-\frac{6}{\gamma^{2}}\sqrt{\frac{8\pi
G}{3}}\int d\phi V^{\prime}\sqrt{V}\exp(-24\pi
G\int^{\phi}d\phi^{\prime} (V^{\prime})^{-1}V\Big).
\end{array}\end{equation}

For the potential in Equation (14)
\begin{equation}
P_{\infty}=\vert \phi^{2}-\frac{\mu^{2}}{g}\vert
\vert\phi\vert^{\frac{3\pi G\mu^{2}}{g}}\exp(-\frac{3}{2}\pi
G\phi^{2})\exp(\frac{-F(\phi)}{\gamma^{2}}),\end{equation} where
\begin{equation}
F(\phi)=\int ^{\phi}U(\phi)\end{equation}and \begin{displaymath}
U(\phi)=6\sqrt{\frac{8\pi G}{3}} V^{\prime}\sqrt{V}\exp\Big(-24\pi
G\int^{\phi}d\phi^{\prime} (V^{\prime})^{-1}V\Big).
\end{displaymath}

We can calculate now the expectation value of $V$ (the
cosmological term $\Lambda$)
\begin{equation}\Lambda=\langle V\rangle=\Big(\int P_{\infty}\Big)^{-1}\int P_{\infty}
V(\phi)\end{equation} for a small $\gamma$ by means of the saddle-%
point method. The critical points are determined by
 $F^{\prime}(\phi_{c})=U(\phi_{c})=0)$, i.e.,

\begin{equation}
V^{\prime}\sqrt{V}\exp\Big(-24\pi G\int^{\phi}d\phi^{\prime}
(V^{\prime})^{-1}V\Big)=0.
\end{equation}

For the potential in Equation (14), $\phi_{c}=0$ is the maximum of
$F$. The minima of $F$ are at $\phi_{c}=\pm \frac{\mu}{\sqrt{g}}$.
We shift the variables in the integrals in the nominator and the
denominator in Equation (58)
\begin{equation}
\phi=\phi_{c}+\gamma^{\nu}q
\end{equation}
with a certain $\nu$.$\nu\neq 1$ is chosen in such a way that
$\gamma^{2} $ in the denominator of the exponent in Equation (56)
cancels with $\gamma$-dependent terms in the expansion of $F$ in
$q$. For the potential in Equation (14) $\nu=\frac{2}{3}$.
Expanding Equation (58) in $\gamma$, we obtain
\begin{equation}
\Lambda=\gamma^{\frac{4}{3}}K
\end{equation}
with a certain constant $K(\gamma)>0$ for any $\gamma$. The
functions $F(\phi)$ as well as $U(\phi)$ are non-analytic at
$g\phi^{2}=\mu^{2}$. If we consider the potential in Equation (17)
or more general $(\phi^{2}-\frac{\mu^{2}}{g})^{4n}$ with a natural
number $n$, then Equation (61) still holds true. It is remarkable
that the index $\frac{4}{3}$ appears also in Kolmogorov's theory
of turbulence as a relation between the dissipation scale and
viscosity. For general potentials $V$, if
$U^{\prime}(\phi_{c})\neq 0$, then $\Lambda\simeq \gamma^{2}$. We
obtain  the result that $\Lambda$ is proportional to the diffusion
constant $\gamma^{2}$ in a different model with a diffusive fluid
in the energy-momentum tensor of Einstein equations
\cite{habampa}.

If $\gamma=0$ (the environmental noise is absent), then we obtain
the Starobinsky solution \cite{starobinsky}(discussed also by
Linde \cite{linde}), which in the Stratonovitch interpretation of
the stochastic  Equation (37) takes the form

\begin{equation}
P_{\infty}=V^{-\frac{3}{4}}\exp(\frac{3}{8G^{2}}\frac{1}{V}).
\end{equation}

However, because $V\rightarrow 0$ when $\phi^{2}\rightarrow
\frac{\mu^{2}}{g} $, this is not a stationary distribution because
it is not integrable at $\phi^{2}= \frac{\mu^{2}}{g} $.   The
point $g\phi^{2}=\mu^{2}$ is beyond the slow-roll approximation.
  To restrict Equation (37) at $\gamma=0$ to the inflationary
domain, we would need to cut the range of $\phi$ by imposing the
boundary conditions as in \cite{venn3}. We show here that with the
thermal noise we can get stationary solutions without imposing
boundary conditions.

 With $\gamma\neq 0$ and   the quantum noise, we write
\begin{equation}
\tilde{P}=H^{-1}a^{-\frac{3}{2}}P.
\end{equation}

Then, Equation (53) for $\tilde{P}$ is
\begin{equation}
\frac{\gamma^{2}}{18}H^{-1}a^{-\frac{3}{2}}\partial_{\phi}\tilde{P}+\frac{1}{8\pi^{2}}H^{\frac{3}{2}}\partial_{\phi}(H^{\frac{5}{2}}a^{\frac{3}{2}}\tilde{P})=
-\frac{1}{3}V^{\prime}a^{\frac{3}{2}}\tilde{P}.\end{equation}

Using the formulas for $H$ and   $a$ (Equation (10)), we obtain
\begin{equation}\begin{array}{l}
\ln\tilde{P}=-6\int d\phi H a^{3}
(\gamma^{2}+\frac{9}{4\pi^{2}}H^{5}a^{3})^{-1}\cr
\times\Big(V^{\prime}+\frac{10}{3}G^{2}VV^{\prime} -32\pi
G^{3}V^{3} (V^{\prime})^{-1}\Big).\end{array}
\end{equation}

From Equation (15), we have $a\rightarrow 0$ exponentially for
$\phi\rightarrow \infty$, then $a^{3}H^{5}\rightarrow 0$, and we
get from Equations (63)--(65) a formula coinciding for large
$\phi$ with Equation (55) (the thermal noise is dominating). We
obtain a stationary solution which is integrable at large $\phi$.
It can be seen that there is no integrability problem at $\phi=0$
as well as at $\phi^{2}=\frac{\mu^{2}}{g}$. For a small $\gamma$
and $a^{3}H^{5}\rightarrow 0$, Equation (65)
 is
of the same form as Equation (55).

As the stationary distribution gives a non-zero weight to the
fluctuations around $\phi^{2}=\frac{\mu^{2}}{g}$, we obtain a
non-zero cosmological constant at arbitrary time. When $V(\phi)$
is large at $\phi\simeq 0$,   the cosmological term will be large
at early time when the stochastic process starts near $\phi\simeq
0$, small when $\phi_{t}(\phi)^{2} $ is approaching
$\phi^{2}=\frac{\mu^{2}}{g}$, and at large time it is approaching
the limit in Equation (61), which determines a small $\Lambda$ for
a small thermal dissipation $\gamma$. The diffusion constant
$\gamma^{2}$ is proportional to the temperature at the end of the
radiation era in warm inflation models (this can be the intrinsic
temperature of the de Sitter space).

\section{Summary and Conclusions} \label{sec.7}
The cosmological term can be interpreted as the expectation value
of the energy density in a quantum state describing the universe
evolution. In the early stage of inflation, it is large when the
quantum state of the universe is far from the ground state. At
later stage, the time evolution moves the state closer to the
ground state, leading to the small value of the energy density.
The final value of the cosmological term is different from zero
because of the thermal and quantum fluctuations at the end of
inflation.We   derive  such a behavior of the cosmological term on
the basis of some reasonable assumptions in the model of inflation
governed by a scalar quantum field (inflaton) described by the
double-well potential. The form of the potential is not essential
for the conclusions.
 We could repeat the calculations for any polynomial double-well potential.
 The asymptotic value of the cosmological term is in a unique way
 determined by the stationary distribution.
The fluctuations which determine the asymptotic value of the
cosmological term depend on the asymptotic probability
distribution of the stochastic process governing the inflaton.
When inflation ends,   the thermal noise overcomes the quantum
noise. As a result, the fluctuations (defining the cosmological
constant) are proportional to the thermal diffusion constant, as
happens in most models of the diffusion processes. If at the final
de Sitter stage an equilibrium is achieved, then by the
dissipation--fluctuation theorem the diffusion constant is
proportional to the final temperature. If we had a good candidate
for the initial state of the universe, we could obtain the precise
time evolution of the cosmological term until  its final
stationary value.

We   study the inflaton contribution to the energy-momentum as the
vacuum energy $<V>$ in the long wave limit (consistent with the
slow-roll approximation). The expectation values of the derivative
terms in the energy-momentum at high momenta of quantum field
theory need renormalzation. The renormalized energy-momentum
contributes (to the lhs of Einstein equations) terms quadratic in
the curvature \cite{s1,s2} (it can also give a divergent
cosmological term leading to the cosmological constant problem
discussed in \cite{weinberg}). Such terms modify equations of
motion (can lead to exponential expansion of $a(t)$ \cite{s2}). In
some models of this type,  the quadratic terms can be expressed by
an extra scalar field. We obtain an effective field theory of
Einstein gravity \cite{trot,ketov} similar to the one studied in
this paper. We think that the model      can  accommodate some
quantum effects  (long wave fluctuations in the $\delta
N$-formalism \cite{starobmulti,ven} and short wave fluctuations by
transformation \cite{trot,ketov}) as well as the classical
$\Lambda CDM$ evolution.

\vspace{6pt}



\begin{thebibliography}{99}
\bibitem{mukhanov}Mukhanov
, V.F.; Chibisov, G.V., Quantum fluctuations and a non-singular
Universe, \emph{Journal of Experimental and Theoretical  Physics
Letters} \textbf{1981}, {\em 33}, 532.


\bibitem{flu1} Hawking
, S.W.,The development of irregularities in a single bubble
inflationary Universe, \emph{Phys. Lett. B} \textbf{1982}, {\em
115}, 295.
\bibitem{flu2} Starobinsky, A.A.,Dynamics of phase transition in the new inflationary
universe scenario and generation of perturbations, \emph{Phys.
Lett. B} \textbf{1982}, {\em 117}, 175.
\bibitem{brand}Mukhanov, V.F.; Feldman, H.A.; Brandenberger, R.H.,
Theory of cosmological perturbations,\emph{Phys. Rep.}
\textbf{1992}, {\em 215}, 203.

\bibitem{guth}Guth, A.,Inflationary Universe:a possible solution of the horizon and flatness problems,\emph{Phys. Rev. D} \textbf{1981}, {\em 23}, 347.

\bibitem{guth-pi}Guth, A.; Pi, S-Y.,Quantum mechanics of the scalar field in the new inflationary universe, \emph{Phys. Rev. D} \textbf{1985}, {\em 32}, 1899.

\bibitem{st1} Albrecht, A.; Ferreira, P.; Joyce, M.; Prokopec, T., Inflation and squeezed
quantum states,\emph{Phys. Rev. D} \textbf{1994}, {\em50}, 4807.
\bibitem{st2} Polarski, D.; Starobinsky, A.A.,Semiclassicality and decoherence of cosmological perturbations, \emph{Class. Quant. Grav.} \textbf{1996}, {\em 13}, 377.


\bibitem{col1} Callan, C.G.; Coleman, S.,Fate of the false vacuum.II.First quantum corrections, \emph{Phys. Rev. D} \textbf{1977}, {\em
16}, 1762.
\bibitem{col2} Coleman, S.; de Lucia, F.,Gravitational effects on and of vacuum decay, \emph{Phys. Rev. D} \textbf{1980}, {\em
21}, 3305.
\bibitem{patra} Patrascioiu, A.,Complex time and the Gaussian approximation, \emph{Phys. Rev. D} \textbf{1981}, {\em 24}, 496.

\bibitem{lima}Perico, E.L.D.; Lima, J.A.S.; Basilakos, S.; Sola, J., Complete cosmic history
with a dynamical $\Lambda=\Lambda(H)$ term, \emph{Phys. Rev. D}
\textbf{2013}, {\em 88}, 063531.

\bibitem{review}Overduin, J.M.; Cooperstock, F.I.,Evolution of the scale factor with a variable cosmological term,
\emph{Phys. Rev. D} \textbf{1998}, {\em 58}, 043506.


\bibitem{weinberg} Weinberg, S.,The cosmological constant problem, \emph{Rev. Mod. Phys.} \textbf{1989}, {\em 61}, 1.

\bibitem{bererafang} Berera, A.; Fang, L.-Z.,Thermally induced density perturbations in the inflation era, \emph{Phys. Rev. Lett.} \textbf{1995}, {\em
74}, 1912.

\bibitem{hindu} Bhattacharya, K.; Mohanty, S.; Rangarayan, R.,Temperature of the inflaton and duration of inflation from Wilkinson microwave anisotropy probe data,
\emph{Phys. Rev. Lett.} \textbf{2006}, {\em 96}, 121302.

\bibitem{warmtun} H\"anggi, P.; Talkner, P.; Borkovec, M.,Reaction-rate theory:fifty years after Kramers,
\emph{Rev. Mod. Phys.} \textbf{1990}, {\em 62}, 251.
\bibitem{starobinsky}Starobinsky, A.A., Stochastic de Sitter (inflationary) stage in the early universe, in Current Topics in Field Theory,Quantum Gravity and Strings. In \emph{Lecture Notes in Physics}; Vega, H.J., Sanchez, N., Eds.;
Springer,Berlin, Germany : 1986; Volume 246.
\bibitem{berrerarev}Berera, A.; Moss, I.G.; Ramos, R.O.,Warm inflation and its microphysical basis, \emph{Rep. Progr. Phys.} \textbf{2009}, {\em 72}, 026901.

\bibitem{vilenkin} Vilenkin, A.,Birth of inflationary universes, \emph{Phys. Rev. D} \textbf{1983}, {\em 27}, 2848.

\bibitem{ramos} Ramos, R.O.; da Silva, L.A.,Power spectrum for inflation models with quantum and thermal noises, \emph{J. Cosmol. Astropart. Phys.} \textbf{2013}, {\em 2013}, 032
\bibitem{int} Haba, Z., Stabilization of Starobinsky-Vilenkin stochastic inflation by an environmental noise, \emph{Int. J. Mod. Phys. D} \textbf{2019}, {\em 28}, 1950085.
\bibitem{habathermal} Haba, Z.,Stochastic inflation with quantum and thermal noise,
 \emph{Eur. Phys. J. C} \textbf{2018}, {\em 78}, 596.
\bibitem{hepj} Haba, Z.,Slow-roll versus stochastic slow-roll inflation \emph{Eur. Phys. J. C} \textbf{2019}, {\em 79}, 906.
\bibitem{starobmulti} Starobinsky, A.A., Multicomponent de Sitter (inflationary) stages and the generation of perturbations \emph{Journal of Experimental and Theoretical Physics Letters} \textbf{1985}, {\em 42}, 152.


\bibitem{ven} Vennin, V.; Starobinsky, A.A.,Correlation functions in stochastic inflation, \emph{Eur. Phys. J. C} \textbf{2015}, {\em 75}, 413.
\bibitem{birrel} Birrel, N.D.; Davies, P.C.W. \emph{Quantum Fields in Curved Space}; Cambridge University Press, Cambridge,England 
: 1982

\bibitem{s1} Mamaev
, S.G.; Mostepanenko, V.M.,Isotropic cosmological models
determined by vacuum quantum effects,  {\em 78}, 20. \emph{Sov.
Phys. JETP} \textbf{1980}, {\em 51}, 9.

\bibitem{s2} Starobinsky, A.A., A new type of isotropic cosmological models without singularity, \emph{Phys. Lett. B} \textbf{1980}, {\em 91}, 99.

\bibitem{starobquan}Starobinsky, A.A.; Yokoyama, J., Equilibrium state of self-interacting
scalar field in the de Sitter background, \emph{Phys. Rev. D}
\textbf{1994}, {\em 50}, 6357.
\bibitem{ikeda} Ikeda, N.; Watanabe, S. \emph{Stochastic
Differential Equations and Diffusion Processes}; North
Holland,Amsterdamm, Netherlands. : 1981

\bibitem{berrera}Berera, A.,Thermal properties of an inflationary universe \emph{Phys. Rev. D} \textbf{1996}, {\em 54},2519.
\bibitem{habaadv} Haba, Z.,Statistical physics of the inflaton decaying in an inhomogeneous
random environment, \emph{Adv. High. Energy. Phys.} \textbf{2018},
{\em 2018}, 7204952.




\bibitem{venn3} Assadullahi, H.; Firouzjahi, H.; Noorbala, M.; Vennin, V.; Wands, D. , Multiple fields in stochastic inflation,\textbf{2016}, {\em2016}, 043 \emph{J. Cosmol. Astropart. Phys.}
\bibitem{risken} Risken, H. \emph{The Fokker-Planck Equation}; Berlin, Germany,Springer add publisher.
, 1989
\bibitem{bard} Bardeeen, J.M.; Steinhardt, P.J.; Turner, M.S.,Spontaneous creation of almost scale-free density perturbations in an inflationary universe,
\emph{Phys. Rev. D} \textbf{1983}, {\em 28}, 679.
\bibitem{ring}
Martin, J.; Ringeval, C.; Vennin, V. Encyclopedia
 Inflationaris. arXiv
:1303.3787
\bibitem{trot}Kehagias, A.; Dizgah, A.M.; Riotto, A.,Remarks on the Starobinsky model of inflation and its descendants,
\emph{Phys. Rev. D} \textbf{2014}, {\em 89}, 043527.
\bibitem{ketov}Ketov, S.V.; Starobinsky, A.A., Inflation and non-minimal scalar curvature coupling in gravity and supergravity, \emph{J. Cosmol. Astropart. Phys.} \textbf{2012}, {\em2012}, 022.

\bibitem{venn}Vennin, V.; Assadullahi, H.; Firouzjahi, H.; Noorbala, M.; Wands, D.,Critical number of fields in stochastic inflation, \emph{Phys. Rev. Lett.} \textbf{2017}, {\em 118}, 031301.

\bibitem{nelson}Nelson, E. \emph{Dynamical Theories of Brownian Motion};Princeton University Press,Princeton,New Jersey, : 1967


\bibitem{krauss}Krauss, L.M.; Dent, J.,  Late time behavior of false vacuum decay:possible
implications for cosmology and metastable inflating states,
\emph{Phys. Rev. Lett.} \textbf{2008}, {\em 100}, 171301.
\bibitem{ssu} Stachowski, A.; Szydlowski, M.; Urbanowski, K. , Quantum mechanical look at
the radioactive-like decay of metastable dark energy \emph{Eur.
Phys. J. C} \textbf{2017}, {\em 77}, 357.
\bibitem{grad}Gradshtein, I.S.; Ryzhik, I.M. \emph{Tables of
Initegrals, Series and Products}; Nauka: Moscow, Russia, 1971. (in
Russian)



\bibitem{ruell}Eckmann, J.-P.; Ruelle, D.,Ergodic theory of chaos and strange attractors, \emph{Rev. Mod. Phys.} \textbf{1985}, {\em
57}, 617.

\bibitem{habampa} Haba, Z., A relation between diffusion, temperature and the cosmological constant, \emph{Mod. Phys. Lett. A} \textbf{2016}, {\em 31}, 1650146.

\bibitem{linde} Linde, A.D., Quantum creation of an open inflationary universe \emph{Phys. Rev. D58},083514 \textbf{1998},





 \end{thebibliography}
\end{document}